\documentclass{rspublic}

\usepackage{longtable}
\usepackage{natbib}
\usepackage{graphicx}
\usepackage{subfig}

\graphicspath{{figs/}}

\begin{document} 
 
\title[Heating of the Solar Atmosphere]{Is Magnetic Topology Important for Heating the Solar Atmosphere?}
 
\author[C.E.~Parnell et al.]{Clare E.~Parnell$^1$, Julie E.H. Stevenson$^1$, James Threlfall$^1$ and Sarah J. Edwards$^{1,2}$} 
 
\affiliation{$^1$School of Mathematics \& Statistics, University of St Andrews, North Haugh, \\ St Andrews, KY16 9SS, UK \\ $^2$School of Mathematics, University of Durham, Durham, UK} 
 
\maketitle 
 
\begin{abstract}{Solar Atmosphere, Corona, Magnetic Fields, Coronal Heating} 
Magnetic fields permeate the entire solar atmosphere weaving an extremely complex pattern on both local and global scales. In order to understand the nature of this tangled web of magnetic fields, its magnetic skeleton, which forms the boundaries between topologically-distinct flux domains, may be determined. The magnetic skeleton consists of null points, separatrix surfaces, spines and separators. The skeleton is often used to clearly visualise key elements of the magnetic configuration, but parts of the skeleton are also locations where currents and waves may collect and dissipate.

In this review, the nature of the magnetic skeleton on both global and local scales, over solar cycle time scales, is explained. The behaviour of wave pulses in the vicinity of both nulls and separators are discussed and so too is the formation of current layers and reconnection at the same features. Each of these processes leads to heating of the solar atmosphere, but collectively do they provide enough heat, spread over a wide enough area, to explain the energy losses throughout the solar atmosphere? Here, we consider this question for the three different solar regions: active regions, open-field regions and the quiet-Sun. 

We find that the heating of active regions and open-field regions are highly unlikely to be due to reconnection or wave dissipation at topological features, but it is possible that these may play a role in the heating of the quiet-Sun. In active regions, the absence of a complex topology may play an important role in allowing large energies to build up and then, subsequently, be explosively released in the form of a solar flare. Additionally, knowledge of the intricate boundaries of open-field regions (which the magnetic skeleton provides) could be very important in determining the main acceleration mechanism(s) of the solar wind.  
\end{abstract} 
 
\section{Introduction}
The question of how the solar corona (or indeed any stellar object with a hot corona)  may be heated has been considered for many decades \cite[see, e.g.][for a recent review]{parnelldemoortel12}, but still remains unanswered. One significant advance has been the recognition that explaining simply how the corona alone is heated is insufficient; the real question is how does the whole solar atmospheric system (the photosphere, chromosphere, transition region and corona) interact and interlink in order to sustain a hot corona? Throughout this volume there are many articles that look to address different aspects of this problem. In this article, we focus on magnetic topology and the role it plays (if any) in the heating of the solar atmosphere.  

Magnetic topology incorporates all properties of a magnetic field that are preserved by ideal displacements. For instance, examples of topological features are: the linkage and knottedness of field lines, null points, their associated separatrix surfaces and spines, as well as separators (see Sec.~\ref{sec:basic-top}). The topology of a magnetic field is not changed by stretching it; it can only change through the process of reconnection. Indeed, a change in topology implies reconnection. 

Almost a decade ago a review article was published on aspects of magnetic topology \citep{longcope05}. Since then, there have been many developments on two fronts: (i) our understanding of the nature of the topology of both global and local coronal magnetic fields and (ii) the role particular topological features play in processes such as magnetic reconnection or wave dissipation. Before reviewing these developments, the basic elements of 3D magnetic topology are outlined in Sec.~\ref{sec:basic-top}. Then, in Sec.~\ref{sec:3Dglobtop}, we investigate the nature of the magnetic skeleton of the global coronal magnetic field, and how it varies over the solar cycle, as well as consider how the magnetic skeleton may evolve under magnetohydrodynamic (MHD) conditions. 

The energy required to heat the solar atmosphere is injected through the photosphere from the convection zone below. This energy arrives in two main forms: rapid motions which incite waves to travel up into the atmosphere or slow motions that cause a gradual stressing of the magnetic field. In Sec.~\ref{sec:photodriving}, we discuss the impact these different driving motions have on the elements of the magnetic skeleton.
Finally, in Sec.~\ref{sec:conc}, we discuss what the consequences of these results are for heating the solar atmosphere and, thus, address the question: is magnetic topology important for coronal heating?   

\section{Basic Elements of the 3D Magnetic Skeleton\label{sec:basic-top}}
\begin{figure}[ht]
\centering
\scalebox{.5}{\includegraphics{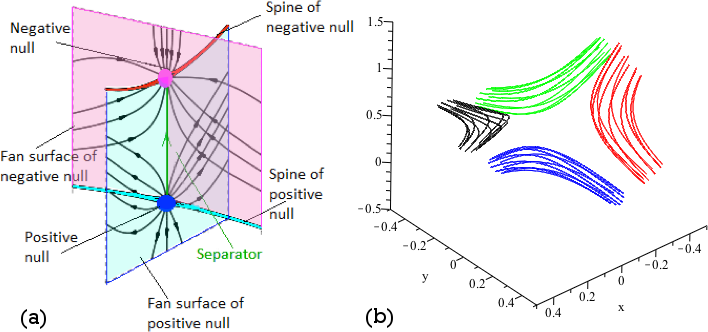}}
\caption{A comparison of the same magnetic field, but viewed in two different ways by: (a) drawing the topological features in the field and (b) drawing sample field lines in four different colours. In (a) the topological features include: a pair of oppositely-signed null points (pink/blue spheres), their associated spines (thick lines) and separatrix surfaces (pink/blue field lines and shaded surfaces). These two separatrix surfaces intersect along a line called a separator (green line) that links the two oppositely-signed null points.}
\label{fig:basic_null_sep_top}
\end{figure}
In order to determine the complexity of a magnetic field in 3D, it is beneficial to determine its magnetic skeleton; this comprises of several features. Magnetic {\it null points} are locations where all three components of the magnetic field equal zero. An infinite number of field lines extend from(/to) a null point forming a surface, known as a {\it separatrix surface}, and a pair of lines extend into(/out from) the null, known as {\it spines}. If the field lines in the separatrix surface are directed out of the null, then the spine field lines will be directed inward, and the null is known as a `positive null'. If these field line orientations are reversed, the null is said to be `negative'. For true 3D magnetic null points no other scenarios are possible, as the magnetic field must be divergence-free \citep[e.g.][]{Parnell96}. Both the local magnetic field near the null and field lines in the separatrix surfaces can take on a range of geometries; these were categorised by \cite{Parnell96}. 

Special field lines, called {\it separators}, may link connecting pairs of null points. These can be formed in several different ways, but the only generic type of separator is that formed by the intersection of two separatrix surfaces from opposite-polarity null points. Fig.~\ref{fig:basic_null_sep_top}a illustrates each of these features in a simple magnetic field involving two null points (one positive, the other negative), their associated separatrix surfaces and spines, and a separator linking the two nulls. 
\begin{figure}[ht]
\centering
\includegraphics[width=3.5cm]{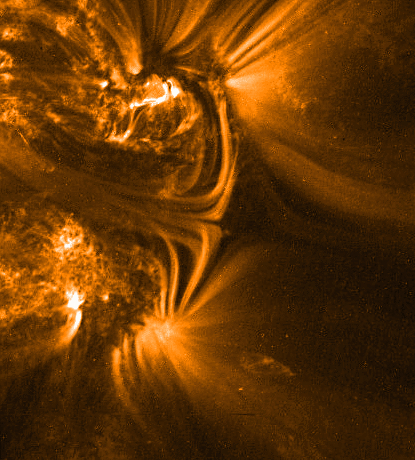}
\caption{Example of a coronal loop structure which has been interpreted as the magnetic field about a (2D) X-type null configuration, but it could equally be the magnetic field about a 3D separator (c.f. Fig.~\ref{fig:basic_null_sep_top}b). This 171\AA\  image, taken by TRACE, shows coronal loops from the active regions NOAA 9149 (above) and 9147 (below), taken at 10:17UT on 4$^{\rm{th}}$ September 2000.}
\label{fig:xtypeobs}
\end{figure}

Fig.~\ref{fig:basic_null_sep_top}b illustrates the same magnetic configuration as that shown in Fig.~\ref{fig:basic_null_sep_top}a, but only for several field lines which do not approach the separator. It is interesting to note that these field lines map out a feature, which, if viewed from above, would look like the field about a 2D X-point. There are a number of observations of coronal loops that may be/are interpreted as X-point like; an early example from TRACE is shown in Fig.~\ref{fig:xtypeobs}. In general, similar coronal loop formations have been attributed to there being a null point at the `X' \citep[e.g.][]{narukage14,freed15}, however, it is possible that a separator may be present instead. 

The local separator magnetic field in planes perpendicular to separators is not always X-type, but may be O-type, if there is a sufficiently large component of current parallel to the separator \citep{parnell10jgr}. In such cases, any magnetic field lines that have undergone separator reconnection (thus having presumably been heated) and so are potentially visible in coronal images, would not show an O-type form. This is because these field lines are actually helical in nature (with a much stronger magnetic field component along the separator than the O-type component perpendicular to it) and, therefore, twist about the separator. So in these cases the expected signature would be a narrow, long, twisted loop-like structure. 

Null points in the corona are, of course, 3D not 2D, but in many instances they may appear 2D, if the field lines in the separatrix surface are strongly aligned along one line \citep[for example, if the local field about the null has a major fan eigenvalue, as explained in][]{Parnell96}. More importantly, null points are local magnetic structures that are identified by determining where ${\bf B}={\bf 0}$. So, by definition, they are single points. Separators, on the other hand, are global magnetic structures and cannot be identified directly by local measurements of the magnetic field. They form lines and, thus, have a far greater spatial extent than null points. In particular, as shown in Sec.~\ref{sec:3Dglobtop}, there are many large-scale separators that arch through the solar atmosphere achieving lengths that are longer than a solar radius. 

In the recent work of \cite{freed15}, the location of coronal null points, identified from low-resolution potential field source surface (PFSS) extrapolations, are compared with AIA observations of the Sun's coronal field. In about 31\% of the cases, \cite{freed15} found that the AIA observations showed structures, such as X-type loop patterns, that could be interpreted as the configuration associated with a null point. However, since separators are field lines that generally extend from or to a null point, it is not unreasonable to imagine that in a proportion of the above cases, the coronal signature may reveal the location of a separator, rather than a null point. It would be interesting to redo this analysis with the locations of the separators also identified.   

\section{3D Coronal Topologies: Global \& Local \label{sec:3Dglobtop}}

All topological features are fundamentally associated with magnetic null points\footnote{In domains where the magnetic field is not closed, separators may instead connect to bald patches \cite[e.g.][]{titov93,titov11}. This type of behaviour occurs where there is a null point lying outside the domain.}. The numbers of coronal null points that exist in potential magnetic field configurations, created by direct extrapolation from observed quiet-Sun regions, have been determined exactly by locating all the nulls \citep[e.g.,][]{regnier08} or estimated using a spectral method \citep[e.g.,][]{longcope09,longcope09b}. Recently, the nulls that occur in PFSS models of the global coronal magnetic field \citep{cook09,platten14,edwards14,freed15,edwards15} have been counted.  

Studies, whose focus is the determination of the number of separators or even the complete magnetic skeleton, are uncommon. \cite{close04} investigated the 
\clearpage
\begin{figure}[H]
\centering
\includegraphics[width=12.5cm]{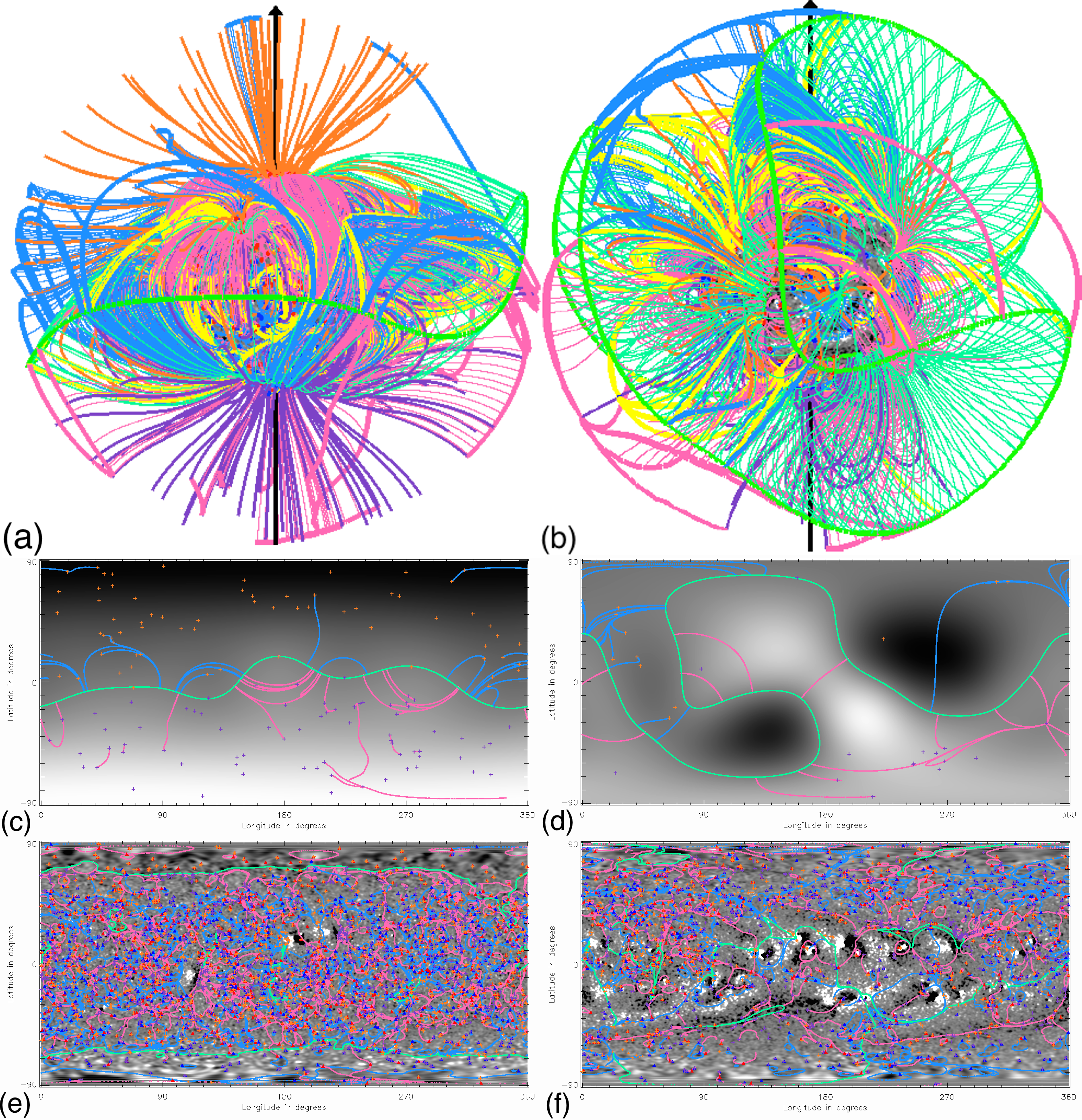}
\caption{The global coronal topology of two PFSS magnetic fields constructed from high-resolution SOLIS synoptic magnetograms taken at (a) solar minimum (CR2083) and at (b) solar maximum (CR2130). The PFSS models use spherical harmonics (with a maximum harmonic number of $l=301$) to extrapolate from the magnetogram data. Topological features marked are: null points (red/blue spheres), spines (orange/purple lines) and separatrix surfaces (pink/blue thin lines). Where the separatrix surfaces intersect the source surface, at $R=2.5R_\odot$, there are thick pink/blue lines. The thick green lines on the source surface indicate the base of the HCS, whilst the thin green lines extending down from these map out the HCS curtains, which divide open/closed field from the Sun. Cuts, at constant radii of (c) \& (d) $R=2.5R_\odot$ and (e) \& (f) $R=1.005R_\odot$ through the topological features, are plotted for the solar minimum case on the left and the solar maximum case on the right. Overplotted on (e) and (f) are all the null points found in each case.}
\label{fig:global_top_min_max}
\end{figure}

\clearpage
\noindent separators found in regions of quiet-Sun, but the most comprehensive survey to date of magnetic skeletons is by \cite{platten14}.

 \subsection{Topological Structures of Global Coronal Fields\label{sec:globtop}}

\cite{platten14} determined the numbers of coronal null points and separators in 496 global coronal potential fields, taken once every Carrington rotation (CR), spanning three solar cycles. The potential fields were constructed from spherical harmonics (with a maximum harmonic number $l=81$) using a PFSS model, where synoptic magnetograms, taken at Kitt Peak by the NSO Vacuum Telescope and SOLIS (low resolution), were used to define the magnetic field on the solar surface and the source surface was taken at $R=2.5R_\odot$. Two example magnetic skeletons of global fields constructed from high-resolution SOLIS magnetographs are shown in Fig.~\ref{fig:global_top_min_max}.  

The study by \cite{platten14} revealed a number of interesting features. (i) Null points above active regions occur at much higher heights than quiet-Sun nulls. (ii) The number of coronal nulls varies out-of-phase with the solar cycle. In the two examples shown in Fig.~\ref{fig:global_top_min_max}, there are 1964 nulls in the solar-minimum case, CR2083, and almost half that number, 1131 nulls, in the solar-maximum case, CR2130. This result is the opposite of that found by \cite{cook09} and \cite{freed15} who find that null-point numbers vary in phase with the solar cycle. However, neither of these studies consider small-scale magnetic features. In \cite{cook09}, simulated (rather than observed) magnetograms are used from a flux transport model in which only active regions (i.e. large-scale structures) are emerged into an otherwise smooth magnetic field. In \cite{freed15}, the global extrapolations are from observed magnetograms and have a maximum harmonic number of just $l=30$, so small-scale structures are missed. This means these two studies can only reliably find nulls at high altitudes, which preferentially occur above active regions. (iii) More null points are found during solar minima with weak polar fields than minima with strong polar fields. 

During solar minima, there are large expanses of quiet-Sun photospheric field that are filled with small-scale magnetic features of both polarities. These produce a complex tangled network of closed coronal field which contains many null points (see Fig.~\ref{fig:global_top_min_max}e). These quiet-Sun fields are surrounded by the dipolar field originating from the Sun's polar regions. If the polar fields are strong then the dipole field will dominate, providing a strong confinement of the quiet-Sun fields and thus limiting quiet-Sun nulls to low heights. On the other hand, if the polar fields are weak, then the dipolar field does not constrain the quiet-Sun field allowing the nulls within it to reside at higher heights above the solar surface. This indicates that during weak polar-field minima (such as the minimum period between cycles 23 and 24) nulls are likely to occur at higher altitudes. Nulls at higher altitudes are more reliably identified, hence, in the work of \cite{platten14} more nulls are found during the weakest polar-field minimum investigated. 
 
 The separators found in global PFSS models come in two forms \citep{platten14}: those that connect pairs of coronal nulls (called null-null separators) and those that connect a coronal null to the ring of nulls that forms the base of the heliospheric current sheet (HCS) on the outer boundary (source surface) of the model. Figs.~\ref{fig:global_top_min_max}c~and~\ref{fig:global_top_min_max}d show cuts at the source surface, through the magnetic skeleton of the two global fields shown in Figs.~\ref{fig:global_top_min_max}a~and~\ref{fig:global_top_min_max}b. The HCS null line is the thick green line in these cuts. There are more null-null separators found during solar minimum than during solar maximum. For instance, 1997 separators are found in CR2083 (solar minimum) of which 1946 are null-null and 51 are null-HCS and just 808 separators, in CR2130 (solar maximum) of which 765 are null-null and 43 null-HCS. The number of null-HCS separators is not strongly dependent on the solar cycle.
 
 Finally, we note that not only do separatrix surfaces emanate from individual coronal nulls, but they also extend out from the HCS null line (green lines in Figs.~\ref{fig:global_top_min_max}a,~\ref{fig:global_top_min_max}b,~\ref{fig:global_top_min_max}e~and~\ref{fig:global_top_min_max}f) separating regions of open field from regions of closed field \citep[e.g.][]{titov11}. At solar minimum, the HCS null line is approximately equatorial, however, at solar maximum, it is highly warped and may cross the poles (Fig.~\ref{fig:global_top_min_max}d). Furthermore, it can split into two or more disconnected loops \citep[e.g.][]{wang14,edwards15a}.   
 
The HCS null line arises because the magnetic field outside the source surface is assumed to be purely radial: where the radial field changes direction a null line forms. In reality, the magnetic field of the solar atmosphere does not all become purely radial at a single radius above the surface and so a null line will never occur in practice. Instead, additional coronal nulls will occur with their associated separatrix surfaces. This means that in reality there will be no null-HCS separators, just additional null-null separators connecting nulls low down in the solar atmosphere to nulls lying above $2.5 R_\odot$. It is not immediately clear how many null-HCS separators currently found will become null-null separators, as opposed to simply disappearing, nor is it clear whether additional null-null separators will be created between nulls above $2.5 R_\odot$ and nulls below $2.5 R_\odot$ that currently do not have null-HCS separators. Further work is required to understand the nature of far coronal magnetic field.
 
 \subsection{Topological Structures in MHD Experiments\label{sec:mhdtop}}

Studies of magnetic topology and null point numbers typically focus on potential magnetic fields. However, there have been a small number of studies investigating the magnetic skeletons of magnetic fields created in highly-dynamic numerical MHD experiments \citep{Haynes2007,maclean09,parnell10apj,mactaggart14}. By highly dynamic, we mean that the magnetic fields undergo significant changes, rather than they evolve at a particularly fast rate. 

\begin{figure}[ht]
\centering
\scalebox{.25}{\includegraphics{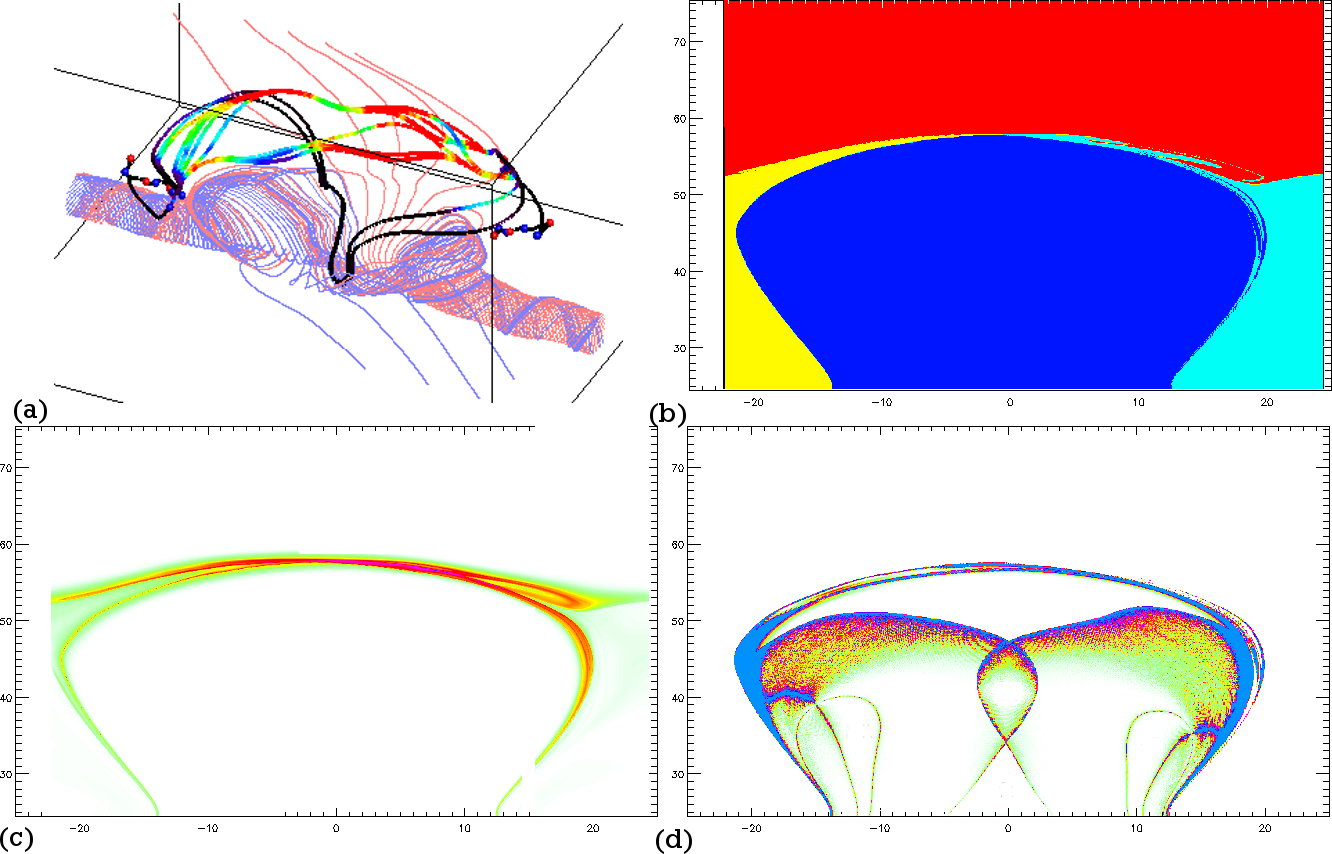}}
\caption{(a) Field lines and elements of the magnetic skeleton in a frame from a flux-emergence experiment \citep{parnell10apj}: pink/blue lines forming the flux rope with some also connecting to the overlying field. The null points (red/blue spheres) and separators (coloured according to the parallel electric field (low - black, high - red) along them) are also plotted. (b) Connectivity of the magnetic field threading a vertical plane perpendicular to the flux rope half-way along its length: overlying field (red), flux rope (blue), flux rope-overlying (yellow), overlying to flux rope (cyan). (c) Integral of the electric field parallel to the magnetic field threading the same plane (white - no reconnection and red/purple - greatest amount of reconnection). (d) Regions of high Q mapped onto the same plane (white - Q = 0 and blue - high Q). Figure based on images from \cite{RestantePhD2011}.}
\label{fig:FE_con_q_intE}
\end{figure}

The experiment considered by \cite{maclean09} and \cite{parnell10apj} involved the emergence of a twisted flux rope from below the solar surface up into an overlying horizontal coronal magnetic field angled at 135$^o$ to the axis of the flux tube \citep{galsgaard07}. In the experiment, the centre of the flux tube was made buoyant such that it rose up and interacted with the overlying field. Initially there were no nulls within the system, but the interaction between the flux tube and coronal field led to the creation of two null clusters, each containing multiple null points, either side of the emerged flux tube \citep{maclean09}. The null points within each null cluster were linked together by single short separators, like beads on a string. All but four nulls were short lived, with the associated cluster separators being equally short lived as the nulls they connect. The long-lived nulls (two of the same sign inside each null cluster) were connected by a multitude of long separators (up to several hundred) that link from one null cluster to the other arching up over the emerged flux tube (Fig.~\ref{fig:FE_con_q_intE}a). Each separator lies at the interface between four topological domains containing: overlying field, flux-tube field and field that connects from the flux tube to the overlying field and vice-versa. These long separators, called intercluster separators, typically survived much longer than the short-lived nulls and cluster separators, but only a few had lifetimes comparable to that of the long lived nulls. 
 
\cite{RestantePhD2011} undertook a thorough analysis of the topology, quasi-separatrix layers \cite[QSLs, see e.g.][]{priest95,demoulin96,titov02,titov07}, and the primary sites of reconnection in this 3D numerical experiment. QSLs are geometrical rather than topological features of the magnetic field. Neighbouring field lines within a QSL will have foot points that are far apart. The extent of the separation of the foot points is measured by the squashing factor, $Q$, which is designed, such that in a given plane, each end of the same field line has the same value of $Q$ \citep{titov02}. Regions of high $Q$ have been associated with reconnection \citep[e.g.][]{titov03,galsgaard03,aulanier05,aulanier06}. 

 Several field lines (seen in pink/blue in Fig.~\ref{fig:FE_con_q_intE}a) form a twisted flux rope that, at the time of this frame, had emerged and partially reconnected with the overlying field. This is illustrated in Fig.~\ref{fig:FE_con_q_intE}a by the few flux-tube field lines that are open and extend up into the solar atmosphere above. Also seen in Fig.~\ref{fig:FE_con_q_intE}a are the null points (red/blue spheres) in the two null clusters and the separators connecting the nulls within the same null cluster (thick black lines) and connecting nulls in different null clusters (thick multicoloured lines).

To compare the magnetic skeleton, QSLs and sites of reconnection specific quantities were calculated on the plane that cuts through the flux tube at right angles, half-way along its length: this plane was chosen because it intersects all the field lines in the volume. 
Separatrix surfaces denote the boundaries between regions of different connectivity. In Fig.~\ref{fig:FE_con_q_intE}b, the intersection points of field lines with this plane are coloured one of four colours according to their connectivity: overlying field (red), flux rope (blue), flux rope-overlying (yellow), overlying-flux rope (cyan). The sites of reconnection (Fig.~\ref{fig:FE_con_q_intE}c) were identified by plotting contours of the integral along a field line of the electric field parallel to the field line \citep[$\int E_{||} dl$ - a non-zero value of this integral is a necessary and sufficient condition for 3D reconnection,][]{schindler88}. Finally, in Fig.~\ref{fig:FE_con_q_intE}d, the value of $Q$ on each field line is indicated, to identify sites where field lines, that start out close to one another, diverge away significantly from each other. Note, to determine Q, both ends of the field line must intersect the photosphere of the model. So, Q is undefined for all overlying field lines, or field lines connected, at one end or the other, to the overlying field. 

Comparing Figs.~\ref{fig:FE_con_q_intE}b,~\ref{fig:FE_con_q_intE}c~and~\ref{fig:FE_con_q_intE}d, it is clear that the sites of reconnection coincide well with boundaries between different connectivities, i.e., they coincide with topological features. However, even though there are regions of high $Q$ in these locations, there are also several other large regions of high $Q$. In these additional high-$Q$ regions, there is no reconnection. By a careful analysis of the magnetic field in these regions, \cite{RestantePhD2011} found that they coincided with significant changes in field line geometry. Therefore, it is important to be cautious about the interpretation of QSLs. 

Not all QSLs are sites of reconnection. But it is also the case that elements in the magnetic skeleton are not always sites of reconnection. In order for reconnection to occur, both a favourable magnetic field configuration and favourable plasma conditions are required to create the non-zero parallel electric field that is essential for 3D reconnection. A key question, therefore, is: Can magnetic field conditions favourable for reconnection arise in the absence of either QSLs or elements of the magnetic skeleton?

Firstly, from a practical point of view, in order to determine $Q$ on a field line, both ends must pass through the surface upon which the separation of originally nearby footprints is measured. This is not always the case, as seen in \cite{RestantePhD2011} where $Q$ was only defined for field lines that crossed the photosphere at both ends, i.e., $Q$ was found for flux rope field lines, but was undefined for flux rope-overlying, overlying-flux rope or purely overlying field lines. Some of these field lines, for which $Q$ could not be defined, are associated with reconnection (as evidenced by the wings in Fig.~\ref{fig:FE_con_q_intE}c). In this experiment, this reconnection is associated with topological features, but it is possible that, in a different situation where there are no topological features, similar field lines where $Q$ can not be defined may undergo reconnection.

Secondly, there are many reconnection experiments, which have not been analysed to determine one or all of the QSLs, the magnetic skeleton or the $\int E_{||} dl$. Therefore, it is not possible to say where exactly the reconnection is occurring and what it is associated with. In one reconnection experiment by \cite{wilmot-smith09}, a very detailed analysis was undertaken. They found that although $Q$ and $\int E_{||} dl$ both had a similar form, the regions of highest $Q$ and highest $\int E_{||} dl$ did not coincide. Further studies of reconnection experiments undertaking similar detailed analysis are required before the exact relationship between $Q$ and $\int E_{||} dl$ may be properly understood.

\section{Responses to Photospheric Driving\label{sec:photodriving}}
It is well known that the source of energy that heats the solar atmosphere and powers events, large and small, from solar flares and coronal mass ejections down to X-ray bright points and nanoflares, comes from convective motions below the solar surface, which either inject energy during the process of flux emergence, or inject energy by driving photospheric magnetic footpoints. The footpoints may be driven at two different rates. Motions that are faster than the Alfv{\'e}n speed generate waves which may propagate into the solar atmosphere, whereas motions that are slower than the Alfv{\'e}n speed cause a stressing of the magnetic field, throughout the atmosphere, that can lead to the generation of current layers. 
Below, we focus on what happens at elements of the magnetic skeleton in response to these motions.

\subsection{Behaviour of Wave Pulses at Topological Features\label{sec:waves}}

In recent years, the advent of telescopes with both high spatial and temporal resolution, as well as high sensitivity,  such as SoHO/EIT, Hinode/XRT, Hinode/EIS, SDO/AIA and CoMP, have enabled a vast number of wave observations to be made.
What happens to waves when they come across features such as nulls, separators or separatrix surfaces? What happens to these topological features when they are hit by waves? 

Firstly, we note that the waves which travel across the magnetic field are fast-magnetoacoustic waves. This type of wave travels at a speed, $c_f = \sqrt{v_A^2+c_s^2}$, where $v_A = B/\sqrt{\mu \rho}$ is the Alfv{\'e}n speed ($B$ is the magnitude of the magnetic field, $\rho$ is the plasma density and $\mu$ is the permeability of free space) and $c_s = \sqrt{p/\rho}$ is the sound speed ($p$ is the plasma pressure). 

The solar corona is a low plasma-beta environment, where the magnetic pressure dominates over the plasma pressure, so $v_A\gg c_s$; thus, $c_f\approx v_A$. 
At a coronal null point, $B = 0$, therefore $v_A=0$ suggesting that a fast-mode wave will be unable to propagate across a null. 

\cite{mclaughlin04} explored the behaviour of fast-wave pulses in the vicinity of a 2D null, in a zero-beta plasma, and showed that the wave essentially becomes trapped. However, the speed of the wave also depends on the sound speed. At a coronal null point $c_f = c_s$, which, although small, is non zero and, hence, a fast-mode wave never actually gets trapped at the null. Instead, the wave undergoes mode coupling. This is investigated in the MHD regime by \cite{mclaughlin06} and \cite{thurgood12} and in the Hall MHD regime by \cite{threlfall12}.  In all cases, the trapping or mode coupling of the fast-mode wave gives rise to localised oscillatory reconnection at the null itself, as well as shocks. Both of these cause heating along the field lines that pass close to the null. 

 Recently, \cite{thurgood13a,thurgood13b}, have studied the behaviour of Alfv\'en waves (which travel at the Alfv{\'e}n speed parallel to the field), in the vicinity of both 2D and 3D nulls. Their work suggests that these waves may become phase-mixed near null points, again causing heating. 
 
As discussed above, both Alfv{\'e}n and fast-magnetoacoustic waves can be affected by the presence of null points \cite[for a review see][]{mclaughlin11}. But what happens to waves as they approach separators or separatrix surfaces? This question has never been addressed before, but since the magnetic field does not reduce in strength near a separator or separatrix surface, it is quite possible that these waves simply cross these features without any change in behaviour. It would still, though, be interesting to investigate what, if any, modification in behaviour is found across separators and separatrix surfaces to confirm or deny the above hypothesis.

In this section, we have so far addressed the behaviour of just single wave pulses interacting with null points. However, analytical and numerical modelling has shown us that single pulses in MHD \citep{hood02} and in Hall MHD \citep{threlfallPhD12} are often limiting cases of full wave trains. So these works may be seen, in some sense, as special cases of the situations we discuss below that undergo continuous driving on the boundary.

\subsection{Effects of Boundary Driving on Topological Features\label{sec:bdrive}}

The work discussed above considers what happens when a single wave pulse arrives at a null. How do topological features respond to continuous stressing as opposed to a single wave? In situations where the local field about the null is driven slowly away from its initial potential state, waves are created that lead to the collapse of the null, creating a localised current layer and triggering reconnection. An enormous body of work exists studying reconnection at 2D null points triggered by boundary driving \citep[e.g. see reviews by][and enclosed references]{priest2000,biskamp2000}. Similarly, reconnection at 3D null points has, in recent years, received quite a bit of attention. 

Due to the additional complexity that the third dimension brings, in 3D, the specifics of the boundary perturbation can cause a variety of different current accumulations, that are associated with different types of reconnection \citep{priest09}. In particular, a 3D potential null point's spine is orthogonal to its separatrix surface. If the angle between the separatrix surface and spine is decreased then this can precipitate a collapse of the null, in the same manner as a 2D null collapses, with a component of current created perpendicular to the plane of the collapse. This causes spine-fan reconnection. If, instead, the separatrix surface is perturbed in such a way that the spine and separatrix surface remain perpendicular, but the field about the spine or about the plane of the separatrix surface is twisted, then instead a current component parallel to the spine is formed. Here, either torsional spine or torsional fan reconnection occurs. The reconnection triggered is different in all three cases. A comprehensive review of 3D null-point reconnection can be found in \cite{priest09} and \cite{pontin13}. Examples of solar scenarios in which reconnection has been triggered at null points by boundary driving includes work by \cite{aulanier00,pariat09,masson09} and \cite{masson14}, but these cases are all aimed at modelling coronal mass ejections (CMEs) or eruptive events, such as solar flares, rather than studying coronal heating.    

The response of magnetic separators and separatrix surfaces to similar boundary driving has received less attention than that for null points (probably because separators and separatrix surfaces are harder to find). However, \cite{haynes07,parnell08,parnell10jgr} and \cite{parnell10apj} study the evolution of the whole magnetic skeleton in driven MHD experiments. \cite{haynes07} showed that boundary driving leads to the build up of currents about the separator and that the resulting reconnection triggered in the current layer naturally leads to the creation of multiple separators connecting the same pair of nulls. \cite{parnell10apj} also found this behaviour in a completely different system. In both cases, the increase in complexity was required to enable a rapid rate of reconnection, and thus, was also connected with the most intense period of heating. As the reconnection rate decreases, so too does the complexity of the magnetic skeleton. This enhancement in complexity/mixing appears to be a generic feature of fast 3D reconnection and has been found in other MHD reconnection experiments, in the absence of null points \citep{pontin11}, as well as at null points \citep{wyper14} and also in a kinetic reconnection experiment \citep{daughton14}.

\subsection{Energy Partitioning due to Reconnection at Topological Features\label{sec:clayers}}

In the above studies, reconnection was triggered by boundary driving. Here, we consider the studies that have taken a different approach and start instead from force-free and non-force-free MHS equilibria involving current layers where reconnection is initiated by micro-instabilities (in MHD experiments, an anomalous reconnection is applied to mimic the triggering of reconnection by micro-instabilities). 

Large Lorentz forces are created within the strong current layers found in non-force-free equilibria. These are counter-balanced by gradients in pressure, providing overall force balance.\footnote{Null points and separators undergo an infinite time collapse, therefore, it is not possible to reach a perfect force balance about such features. Instead, a state is reached in which the total force is zero everywhere, except at the nulls and separators themselves, where it is very small \citep[e.g.][]{Fuentes-Fernandez12c,Fuentes-Fernandez13,Stevenson15}.} 

One interesting aspect of these reconnection studies is the ability to follow, in great detail, the energy partitioning and transport resulting from the reconnection. \cite{longcope07} and \cite{longcope12} considered what happens following reconnection in a force-free equilibria with a current sheet located at a 2D null point. They analytically determined that the proportion of magnetic energy going directly into Ohmic heating was much smaller than that converted into kinetic energy and ultimately released by viscous dissipation. Thus, they found that 2D magnetic reconnection (in an initially force-free system) could cause significant heating over a wide region, potentially far from the null point.

On the other-hand, investigations of the energy partitioning resulting from reconnection in non-force-free MHS equilibria with a current sheet at a 2D null point by \cite{Fuentes-Fernandez12c} and \cite{Fuentes-Fernandez13} find that the kinetic energy created in the system is an order of magnitude less than the changes in magnetic and internal energy. In these studies, Ohmic heating, directly in the vicinity of the null point, dominates over all other forms of heating. 

Both high and low plasma-beta cases have been considered\footnote{Exactly at the magnetic null point the plasma beta will be infinite no matter what value the plasma pressure is there.}, with more energy going into the waves in the low-beta case, than the high-beta case. These waves are created by the sudden loss of force balance and emanate out from the edge of the diffusion region. They are magnetoacoustic in nature and travel at the fast-mode speed everywhere except, along the separatrices (which are field lines), where they travel at the slow-mode speed. 

\begin{figure}[ht]
\centering
\scalebox{.3}{\includegraphics{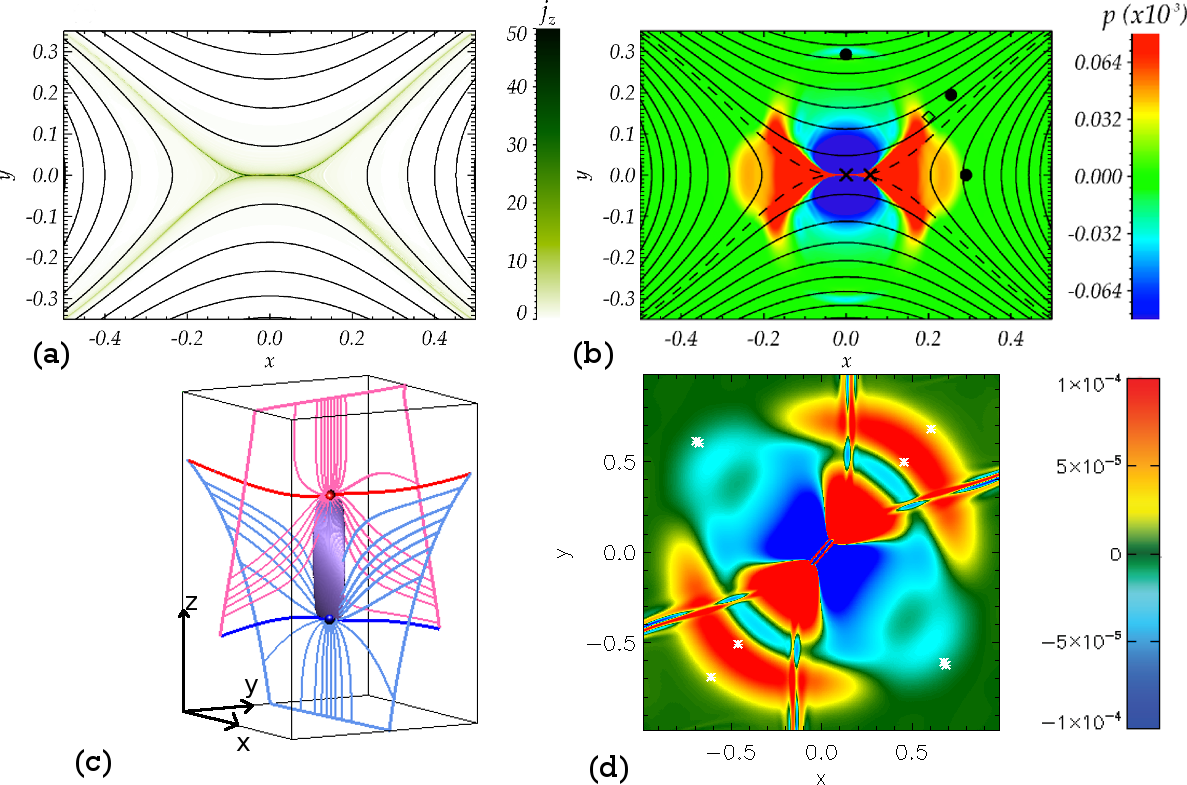}}
\caption{(a) Contour plot of the current in a non-force-free, MHS equilibria about a 2D null point. (b) Perturbation in pressure indicating the waves and flows created by the localised reconnection at this 2D null \citep[created from][]{Fuentes-Fernandez13}. (c) The magnetic skeleton of a non-force-free 3D MHS equilibria involving a separator current layer (indicated by an isosurface). (d) Perturbations in pressure in the $xy$-plane half way along the separator indicating the waves and flows created by reconnection in the current layer \citep[created from][]{Stevenson15b}.}
\label{fig:perturbedp}
\end{figure}
\cite{Stevenson15b} considered a system involving a non-force-free 3D MHS equilibria in which a separator current layer exists with current parallel to the separator. Only high-beta cases have been considered thus far; in all cases the waves and flows generated by the reconnection carry just a fraction of the magnetic energy converted during the reconnection away from the reconnection site. It would be interesting to investigate a low-beta case of separator reconnection, where, there would be no issues about the plasma beta becoming infinite, instead remaining low throughout the reconnection site and surrounding region. 

The reconnection proceeds in essentially the same way as it does at a 2D null point. Rapid reconnection occurs within the separator current layer, converting most of the released magnetic energy directly into internal energy via Ohmic heating at the separator. This causes a loss of force balance which generates waves that are launched from the edges of the diffusion region. In planes perpendicular to the separator, these waves behave in a very similar fashion to that seen in the 2D null point case (Fig.~\ref{fig:perturbedp}). 

In the wake of these waves, the system tries to regain force balance and so flows are generated. These flows drive slow, steady reconnection at the separator, which lasts much longer than the initial fast-reconnection stage. From \cite{Stevenson15b}, it is clear that the separator reconnection does not occur at the 3D null points that lie at either end of the separator, but occurs at some position along its length. The initial seed location of the reconnection is likely to be where the separator current reaches a peak. This could be anywhere along the length of the separator, depending on the nature of the surrounding magnetic field and plasma.    

The partitioning of magnetic energy released by reconnection is still an unanswered question, even in the relatively simple MHD regime where the magnetic energy can only become kinetic or internal energy. In reality, it could also be converted into non-thermal energy producing accelerated particles and thus even less will then go into internal or kinetic energy. In order to explain the solar atmospheric heating problem, it is essential that the energy partitioning due to reconnection is properly understood, whether the reconnection occurs at topological features or in their absence.

\section{Discussion \& Conclusions\label{sec:conc}}

In this paper, we have given a brief review of both the nature of the topology throughout the corona, as well as discussing the different energy release mechanisms associated with the elements of the magnetic skeleton. Here, we consider the question posed in the title: is magnetic topology important for coronal heating? The answer to this is not straightforward and depends on a number of things, (i) how the question is interpreted, (ii) for what location in the solar atmosphere, and for what period during the solar cycle, the question is posed.

Let us first consider whether the trapping, coupling and dissipation of waves at null points or reconnection at null points, separators and on separatrix surfaces is likely to be important for solar atmospheric heating. As already mentioned, several thousand null points and separators have been found in a single PFSS extrapolation made from high-resolution SOLIS data. Using HMI data even more nulls \citep{edwards15} and associated separators are found . At even higher resolution, we are likely to find millions of nulls and separators in the solar atmosphere at any one time. Additionally, due to the highly-dynamic nature of the solar atmosphere, there will be many more topological features than those identified in potential or (non-)linear force-free magnetic fields. However, the majority of these additional nulls will be located in quiet-Sun regions where the surface magnetic field is immensely complex. In active regions, the surface magnetic field is relatively simple, since it is dominated by just a few large-scale sunspots or pores. Thus, the magnetic field above active regions contains considerably fewer topological features than the equivalent sized region of quiet Sun. 

From our current level of understanding, direct heating at topological features is unlikely to be important in active regions, the warmest coronal regions on the Sun. Instead, within large (active-region) scale volumes of the solar atmosphere possible heating mechanisms (in no particular order) are (i) reconnection associated with the complex driving of simple fields \citep[e.g.][]{galsgaard96,bowness13, wilmot-smith10,pontin11}, (ii) reconnection triggered by local instabilities in, for example, twisted flux tubes \citep[e.g.][]{browning08,bareford10,bareford13} and (iii) the dissipation of waves via phase mixing \citep[e.g.][]{heyvaerts83}, resonant absorption \citep[e.g.][]{ionson78} or mode coupling \citep[for a review of wave dissipation mechanisms, see][]{demoortel12}.  

In open-field regions, at present it is very difficult to get reliable measurements of the magnetic fields. Initial results suggest that isolated coronal nulls that are not connected to any separators are fairly common-place in open-field regions \citep{edwards15}. Until there are more reliable polar-field measurements, the numbers of coronal nulls and associated features will remain largely unknown. Thus, it is not possible to speculate on whether reconnection at topological features plays a major role in the acceleration of the fast solar wind. 

In order to produce a wind that emanates out over the entire area of an open-field region, there would have to be an extremely large number of coronal nulls: indeed most probably an unfeasably large number. Therefore wave dissipation without null points remains the most likely heating/acceleration mechanism in coronal holes \citep[for a review, see][]{ofman10}.  

The topology of the quiet-Sun magnetic field is extremely complex. As the resolution and sensitivity of magnetographs increases so too will levels of observable mixed polarity field. This will not continue indefinitely and, although not reached yet, there is likely to be a limit at which increased resolution does not lead to additional small-scale features.
So it is possible (depending upon the size, which is as yet unknown, of the finest scale features) that there are sufficient numbers of topological features to enable heat due to reconnection and wave dissipation mechanisms at topological features to spread over a sufficient area to explain the (apparently uniform) heating of the so-called background corona.  The question then arises: can enough energy be dissipated, by reconnection or waves, at these features? Considerably more investigation and theoretical modelling needs to be undertaken to answer this. At the moment, numerical models cannot replicate the appropriate plasma parameters found on the Sun, so determining the correct energies is difficult. 

The magnetic skeleton associated with the finest-scale photospheric features is likely to be on a scale much larger than that found, for instance, in the case where (through dynamic reconnection) the local structures about a single separator cascade to small-scales producing many hundreds of separators \citep[e.g.][]{parnell10apj}. In such dynamical cascades that approach/reach a turbulent state, the magnetic skeleton may become overly complex and evolve too rapidly for detailed knowledge of each and every null point and separator to be of use. However, knowledge of the skeleton before and after this turbulent behaviour could help to identify the location of where the turbulent reconnection will occur and the consequences of the resulting heating.

The energy built up over even a relatively short period of slow stressing of the field (shearing) and conversion to heat by rapid reconnection (as a result of micro-instabilities) is likely to be greater than any heating due to wave pulses dissipated at topological features. Furthermore, since waves are trapped only at nulls and not at other topological features, the spread of energy through wave dissipation at coronal nulls is unlikely to be as great as that from reconnection at coronal nulls, separators and separatrix surfaces combined. 

It is worth noting here that, although reconnection leads to localised direct Ohmic dissipation, it also launches waves. The energy partitioning of the magnetic energy released during reconnection is still a major unanswered question. As already mentioned, this question has been tackled from an MHD point of view to ascertain whether Ohmic or viscous heating dominates \citep{longcope07,birn09,FuentesFernandez2012a,longcope12,FuentesFernandez2012b,Stevenson15}; there is currently no definitive answer to this question.
Nonetheless, it is clear that waves can drive reconnection and reconnection can launch waves and, therefore, the discussion as to whether waves or reconnection is the key heating mechanism is difficult to disentangle, as these mechanisms are completely interlinked. 
  
The question regarding the importance of topological features as the major heating sites in the solar atmosphere very much depends on the nature of the magnetic field within a given region. What the topological features (such as nulls, separators and, in particular, separatrix surfaces) do is reveal the likely nature of the field. Using global models, the boundaries of open-field regions can be identified. Furthermore, the volumes of closed field within the open-field regions can be identified, enabling a much better understanding of the true expanse of the open-field volume and also the nature of the magnetic fields within the volume. Both the location and geometry of the boundaries  of large-scale closed field regions (such as those associated with active regions) can also be determined, as can the boundaries of the multitude of small-scale closed-field regions that make up the quiet Sun. Thus, in terms of coronal heating, knowledge of the magnetic topology is crucial in order to identify the nature of the solar atmosphere where different atmospheric heating mechanisms are likely to dominate.   

In this article, we have simply considered whether topological features are important for coronal heating. Thus, we have not considered whether other features, such as QSLs, which are principally geometrical in nature, are important for coronal heating. It is certainly clear that reconnection can occur in the absence of nulls and separators \citep[e.g.][]{schindler88,hesse88,galsgaard96,aulanier05,aulanier06,wilmot-smith10,pontin11}, as can the dissipation of waves \citep[e.g.][]{ionson78,heyvaerts83}. Furthermore, we have not addressed the question of whether topological features are important for other solar events, such as solar flares, CMEs, X-ray jets or X-ray bright points. There are at present a number of published works which strongly suggest that they are indeed important \cite[e.g.][]{parnell94,aulanier00,ugarte-urra07,pariat09,masson09}.    

\subsection*{Acknowledgements} 
CEP and JT acknowledge the support of STFC through the St Andrew's SMTG consolidated grant. JEHS is supported by STFC as a PhD student. SJE is supported STFC through the Durham University Impact Acceleration Account.
 
\bibliographystyle{rspublicnat}

\bibliography{references}

\begin{thebibliography}{78}
\providecommand{\natexlab}[1]{#1}
\expandafter\ifx\csname urlstyle\endcsname\relax
  \providecommand{\doi}[1]{doi:\discretionary{}{}{}#1}\else
  \providecommand{\doi}{doi:\discretionary{}{}{}\begingroup
  \urlstyle{rm}\Url}\fi

\bibitem[{{Aulanier} \emph{et~al.}(2000){Aulanier}, {DeLuca}, {Antiochos},
  {McMullen} \& {Golub}}]{aulanier00}
{Aulanier}, G., {DeLuca}, E.~E., {Antiochos}, S.~K., {McMullen}, R.~A. \&
  {Golub}, L. 2000 {The Topology and Evolution of the Bastille Day Flare}.
\newblock \emph{Astrophys. J.}, \textbf{540}, 1126--1142.
\newblock (\doi{10.1086/309376})

\bibitem[{{Aulanier} \emph{et~al.}(2005){Aulanier}, {Pariat} \&
  {D{\'e}moulin}}]{aulanier05}
{Aulanier}, G., {Pariat}, E. \& {D{\'e}moulin}, P. 2005 {Current sheet
  formation in quasi-separatrix layers and hyperbolic flux tubes}.
\newblock \emph{Astron. Astrophys.}, \textbf{444}, 961--976.
\newblock (\doi{10.1051/0004-6361:20053600})

\bibitem[{{Aulanier} \emph{et~al.}(2006){Aulanier}, {Pariat}, {D{\'e}moulin} \&
  {DeVore}}]{aulanier06}
{Aulanier}, G., {Pariat}, E., {D{\'e}moulin}, P. \& {DeVore}, C.~R. 2006
  {Slip-Running Reconnection in Quasi-Separatrix Layers}.
\newblock \emph{Solar Phys.}, \textbf{238}, 347--376.
\newblock (\doi{10.1007/s11207-006-0230-2})

\bibitem[{{Bareford} \emph{et~al.}(2010){Bareford}, {Browning} \& {van der
  Linden}}]{bareford10}
{Bareford}, M.~R., {Browning}, P.~K. \& {van der Linden}, R.~A.~M. 2010 {A
  nanoflare distribution generated by repeated relaxations triggered by kink
  instability}.
\newblock \emph{Astron. Astrophys.}, \textbf{521}, A70.
\newblock (\doi{10.1051/0004-6361/201014067})

\bibitem[{{Bareford} \emph{et~al.}(2013){Bareford}, {Hood} \&
  {Browning}}]{bareford13}
{Bareford}, M.~R., {Hood}, A.~W. \& {Browning}, P.~K. 2013 {Coronal heating by
  the partial relaxation of twisted loops}.
\newblock \emph{Astron. Astrophys.}, \textbf{550}, A40.
\newblock (\doi{10.1051/0004-6361/201219725})

\bibitem[{{Birn} \emph{et~al.}(2009){Birn}, {Fletcher}, {Hesse} \&
  {Neukirch}}]{birn09}
{Birn}, J., {Fletcher}, L., {Hesse}, M. \& {Neukirch}, T. 2009 {Energy Release
  and Transfer in Solar Flares: Simulations of Three-Dimensional Reconnection}.
\newblock \emph{Astrophys. J.}, \textbf{695}, 1151--1162.
\newblock (\doi{10.1088/0004-637X/695/2/1151})

\bibitem[{{Biskamp}(2000)}]{biskamp2000}
{Biskamp}, D. 2000 \emph{{Magnetic Reconnection in Plasmas}}.
\newblock Cambridge University Press, Cambridge, UK.

\bibitem[{{Bowness} \emph{et~al.}(2013){Bowness}, {Hood} \&
  {Parnell}}]{bowness13}
{Bowness}, R., {Hood}, A.~W. \& {Parnell}, C.~E. 2013 {Coronal heating and
  nanoflares: current sheet formation and heating}.
\newblock \emph{Astron. Astrophys.}, \textbf{560}, A89.
\newblock (\doi{10.1051/0004-6361/201116652})

\bibitem[{{Browning} \emph{et~al.}(2008){Browning}, {Gerrard}, {Hood}, {Kevis}
  \& {van der Linden}}]{browning08}
{Browning}, P.~K., {Gerrard}, C., {Hood}, A.~W., {Kevis}, R. \& {van der
  Linden}, R.~A.~M. 2008 {Heating the corona by nanoflares: simulations of
  energy release triggered by a kink instability}.
\newblock \emph{Astron. Astrophys.}, \textbf{485}, 837--848.
\newblock (\doi{10.1051/0004-6361:20079192})

\bibitem[{{Close} \emph{et~al.}(2004){Close}, {Parnell} \& {Priest}}]{close04}
{Close}, R.~M., {Parnell}, C.~E. \& {Priest}, E.~R. 2004 {Separators in 3D
  Quiet-Sun Magnetic Fields}.
\newblock \emph{Solar Phys.}, \textbf{225}, 21--46.
\newblock (\doi{10.1007/s11207-004-3259-0})

\bibitem[{{Cook} \emph{et~al.}(2009){Cook}, {Mackay} \& {Nandy}}]{cook09}
{Cook}, G.~R., {Mackay}, D.~H. \& {Nandy}, D. 2009 {Solar Cycle Variations of
  Coronal Null Points: Implications for the Magnetic Breakout Model of Coronal
  Mass Ejections}.
\newblock \emph{Astrophys. J.}, \textbf{704}, 1021--1035.
\newblock (\doi{10.1088/0004-637X/704/2/1021})

\bibitem[{{Daughton} \emph{et~al.}(2014){Daughton}, {Nakamura}, {Karimabadi},
  {Roytershteyn} \& {Loring}}]{daughton14}
{Daughton}, W., {Nakamura}, T.~K.~M., {Karimabadi}, H., {Roytershteyn}, V. \&
  {Loring}, B. 2014 {Computing the reconnection rate in turbulent kinetic
  layers by using electron mixing to identify topology}.
\newblock \emph{Phys. Plasmas}, \textbf{21}(5), 052307.
\newblock (\doi{10.1063/1.4875730})

\bibitem[{{De Moortel} \& {Nakariakov}(2012)}]{demoortel12}
{De Moortel}, I. \& {Nakariakov}, V.~M. 2012 {Magnetohydrodynamic waves and
  coronal seismology: an overview of recent results}.
\newblock \emph{Roy. Soc. Lond. Phil. Trans. Series A}, \textbf{370},
  3193--3216.
\newblock (\doi{10.1098/rsta.2011.0640})

\bibitem[{{D{\'e}moulin} \emph{et~al.}(1996){D{\'e}moulin}, {Priest} \&
  {Lonie}}]{demoulin96}
{D{\'e}moulin}, P., {Priest}, E.~R. \& {Lonie}, D.~P. 1996 {Three-dimensional
  magnetic reconnection without null points 2. Application to twisted flux
  tubes}.
\newblock \emph{J. Geophys. Res.}, \textbf{101}, 7631--7646.
\newblock (\doi{10.1029/95JA03558})

\bibitem[{{Edwards}(2014)}]{edwards14}
{Edwards}, S.~J. 2014 {On the Topology of Global Coronal Magnetic Fields}.
\newblock Ph.D. thesis, University of St Andrews.

\bibitem[{{Edwards} \& {Parnell}(2015)}]{edwards15}
{Edwards}, S.~J. \& {Parnell}, C.~E. 2015 {Null point distribution in global
  coronal potential field extrapolations}.
\newblock \emph{Solar Phys.}, \textbf{submitted}.

\bibitem[{{Edwards} \emph{et~al.}(2015){Edwards}, {Parnell}, {Harra}, {Culhane}
  \& {Brooks}}]{edwards15a}
{Edwards}, S.~J., {Parnell}, C.~E., {Harra}, L.~K., {Culhane}, J.~L. \&
  {Brooks}, D. 2015 {Modelling active-region open-field regions and comparing
  with remotely sensed upflows}.
\newblock \emph{Solar Phys.}, \textbf{submitted}.

\bibitem[{{Freed} \emph{et~al.}(2015){Freed}, {Longcope} \&
  {McKenzie}}]{freed15}
{Freed}, M.~S., {Longcope}, D.~W. \& {McKenzie}, D.~E. 2015 {Three-Year Global
  Survey of Coronal Null Points from Potential-Field-Source-Surface (PFSS)
  Modeling and Solar Dynamics Observatory (SDO) Observations}.
\newblock \emph{Solar Phys.}, \textbf{290}, 467--490.
\newblock (\doi{10.1007/s11207-014-0616-5})

\bibitem[{{Fuentes-Fern{\'a}ndez} \& {Parnell}(2012)}]{Fuentes-Fernandez12c}
{Fuentes-Fern{\'a}ndez}, J. \& {Parnell}, C.~E. 2012 {Magnetohydrodynamics
  dynamical relaxation of coronal magnetic fields. III. 3D spiral nulls}.
\newblock \emph{Astron. Astrophys.}, \textbf{544}, A77.
\newblock (\doi{10.1051/0004-6361/201219190})

\bibitem[{{Fuentes-Fern{\'a}ndez} \& {Parnell}(2013)}]{Fuentes-Fernandez13}
{Fuentes-Fern{\'a}ndez}, J. \& {Parnell}, C.~E. 2013 {Magnetohydrodynamics
  dynamical relaxation of coronal magnetic fields. IV. 3D tilted nulls}.
\newblock \emph{Astron. Astrophys.}, \textbf{554}, A145.
\newblock (\doi{10.1051/0004-6361/201220346})

\bibitem[{{Fuentes-Fern{\'a}ndez}
  \emph{et~al.}(2012{\natexlab{\emph{a}}}){Fuentes-Fern{\'a}ndez}, {Parnell},
  {Hood}, {Priest} \& {Longcope}}]{FuentesFernandez2012a}
{Fuentes-Fern{\'a}ndez}, J., {Parnell}, C.~E., {Hood}, A.~W., {Priest}, E.~R.
  \& {Longcope}, D.~W. 2012{\natexlab{\emph{a}}} {Consequences of spontaneous
  reconnection at a two-dimensional non-force-free current layer}.
\newblock \emph{Phys. Plasmas}, \textbf{19}(2), 022\,901.
\newblock (\doi{10.1063/1.3683002})

\bibitem[{{Fuentes-Fern{\'a}ndez}
  \emph{et~al.}(2012{\natexlab{\emph{b}}}){Fuentes-Fern{\'a}ndez}, {Parnell} \&
  {Priest}}]{FuentesFernandez2012b}
{Fuentes-Fern{\'a}ndez}, J., {Parnell}, C.~E. \& {Priest}, E.~R.
  2012{\natexlab{\emph{b}}} {The onset of impulsive bursty reconnection at a
  two-dimensional current layer}.
\newblock \emph{Phys. Plasmas}, \textbf{19}(7), 072901.
\newblock (\doi{10.1063/1.4729334})

\bibitem[{{Galsgaard} \emph{et~al.}(2007){Galsgaard}, {Archontis},
  {Moreno-Insertis} \& {Hood}}]{galsgaard07}
{Galsgaard}, K., {Archontis}, V., {Moreno-Insertis}, F. \& {Hood}, A.~W. 2007
  {The Effect of the Relative Orientation between the Coronal Field and New
  Emerging Flux. I. Global Properties}.
\newblock \emph{Astrophys. J.}, \textbf{666}, 516--531.
\newblock (\doi{10.1086/519756})

\bibitem[{{Galsgaard} \& {Nordlund}(1996)}]{galsgaard96}
{Galsgaard}, K. \& {Nordlund}, {\AA}. 1996 {Heating and activity of the solar
  corona 1. Boundary shearing of an initially homogeneous magnetic field}.
\newblock \emph{J. Geophys. Res.}, \textbf{101}, 13\,445--13\,460.
\newblock (\doi{10.1029/96JA00428})

\bibitem[{{Galsgaard} \emph{et~al.}(2003){Galsgaard}, {Titov} \&
  {Neukirch}}]{galsgaard03}
{Galsgaard}, K., {Titov}, V.~S. \& {Neukirch}, T. 2003 {Magnetic Pinching of
  Hyperbolic Flux Tubes. II. Dynamic Numerical Model}.
\newblock \emph{Astrophys. J.}, \textbf{595}, 506--516.
\newblock (\doi{10.1086/377258})

\bibitem[{{Haynes} \emph{et~al.}(2007{\natexlab{\emph{a}}}){Haynes}, {Parnell},
  {Galsgaard} \& {Priest}}]{Haynes2007}
{Haynes}, A.~L., {Parnell}, C.~E., {Galsgaard}, K. \& {Priest}, E.~R.
  2007{\natexlab{\emph{a}}} {Magnetohydrodynamic evolution of magnetic
  skeletons}.
\newblock \emph{Roy. Soc. Lon. Proc. Series A}, \textbf{463}, 1097--1115.
\newblock (\doi{10.1098/rspa.2007.1815})

\bibitem[{{Haynes} \emph{et~al.}(2007{\natexlab{\emph{b}}}){Haynes}, {Parnell},
  {Galsgaard} \& {Priest}}]{haynes07}
{Haynes}, A.~L., {Parnell}, C.~E., {Galsgaard}, K. \& {Priest}, E.~R.
  2007{\natexlab{\emph{b}}} {Magnetohydrodynamic evolution of magnetic
  skeletons}.
\newblock \emph{Roy. Soc. Lond. Proc. Series A}, \textbf{463}, 1097--1115.
\newblock (\doi{10.1098/rspa.2007.1815})

\bibitem[{{Hesse} \& {Schindler}(1988)}]{hesse88}
{Hesse}, M. \& {Schindler}, K. 1988 {A theoretical foundation of general
  magnetic reconnection}.
\newblock \emph{J. Geophys. Res.}, \textbf{93}, 5559--5567.
\newblock (\doi{10.1029/JA093iA06p05559})

\bibitem[{{Heyvaerts} \& {Priest}(1983)}]{heyvaerts83}
{Heyvaerts}, J. \& {Priest}, E.~R. 1983 {Coronal heating by phase-mixed shear
  Alfven waves}.
\newblock \emph{Astron. Astrophys.}, \textbf{117}, 220--234.

\bibitem[{{Hood} \emph{et~al.}(2002){Hood}, {Brooks} \& {Wright}}]{hood02}
{Hood}, A.~W., {Brooks}, S.~J. \& {Wright}, A.~N. 2002 {Coronal heating by the
  phase mixing of individual pulses propagating in coronal holes}.
\newblock \emph{Roy. Soc. Lond. Proc. Series A}, \textbf{458}, 2307.
\newblock (\doi{10.1098/rspa.2002.0959})

\bibitem[{{Ionson}(1978)}]{ionson78}
{Ionson}, J.~A. 1978 {Resonant absorption of Alfvenic surface waves and the
  heating of solar coronal loops}.
\newblock \emph{Astrophys. J.}, \textbf{226}, 650--673.
\newblock (\doi{10.1086/156648})

\bibitem[{{Longcope}(2005)}]{longcope05}
{Longcope}, D. 2005 {Topological Methods for the Analysis of Solar Magnetic
  Fields}.
\newblock \emph{Living Rev. Solar Phys.}, \textbf{7}(2).

\bibitem[{{Longcope} \emph{et~al.}(2009){Longcope}, {Parnell} \&
  {DeForest}}]{longcope09b}
{Longcope}, D., {Parnell}, C. \& {DeForest}, C. 2009 {The Density of Coronal
  Null Points from Hinode and MDI}.
\newblock In \emph{The second hinode science meeting: Beyond discovery-toward
  understanding} (eds B.~{Lites}, M.~{Cheung}, T.~{Magara}, J.~{Mariska} \&
  K.~{Reeves}), vol. 415 of \emph{Astronomical Society of the Pacific
  Conference Series}, p. 178.

\bibitem[{{Longcope} \& {Parnell}(2009)}]{longcope09}
{Longcope}, D.~W. \& {Parnell}, C.~E. 2009 {The Number of Magnetic Null Points
  in the Quiet Sun Corona}.
\newblock \emph{Solar Phys.}, \textbf{254}, 51--75.
\newblock (\doi{10.1007/s11207-008-9281-x})

\bibitem[{{Longcope} \& {Priest}(2007)}]{longcope07}
{Longcope}, D.~W. \& {Priest}, E.~R. 2007 {Fast magnetosonic waves launched by
  transient, current sheet reconnection}.
\newblock \emph{Phys. Plasmas}, \textbf{14}(12), 122905.
\newblock (\doi{10.1063/1.2823023})

\bibitem[{{Longcope} \& {Tarr}(2012)}]{longcope12}
{Longcope}, D.~W. \& {Tarr}, L. 2012 {The Role of Fast Magnetosonic Waves in
  the Release and Conversion via Reconnection of Energy Stored by a Current
  Sheet}.
\newblock \emph{Astrophys. J.}, \textbf{756}, 192.
\newblock (\doi{10.1088/0004-637X/756/2/192})

\bibitem[{{Maclean} \emph{et~al.}(2009){Maclean}, {Parnell} \&
  {Galsgaard}}]{maclean09}
{Maclean}, R.~C., {Parnell}, C.~E. \& {Galsgaard}, K. 2009 {Is Null-Point
  Reconnection Important for Solar Flux Emergence?}
\newblock \emph{Solar Phys.}, \textbf{260}, 299--320.
\newblock (\doi{10.1007/s11207-009-9458-y})

\bibitem[{{MacTaggart} \& {Haynes}(2014)}]{mactaggart14}
{MacTaggart}, D. \& {Haynes}, A.~L. 2014 {On magnetic reconnection and flux
  rope topology in solar flux emergence}.
\newblock \emph{MNRAS}, \textbf{438}, 1500--1506.
\newblock (\doi{10.1093/mnras/stt2285})

\bibitem[{{Masson} \emph{et~al.}(2014){Masson}, {McCauley}, {Golub}, {Reeves}
  \& {DeLuca}}]{masson14}
{Masson}, S., {McCauley}, P., {Golub}, L., {Reeves}, K.~K. \& {DeLuca}, E.~E.
  2014 {Dynamics of the Transition Corona}.
\newblock \emph{Astrophys. J.}, \textbf{787}, 145.
\newblock (\doi{10.1088/0004-637X/787/2/145})

\bibitem[{{Masson} \emph{et~al.}(2009){Masson}, {Pariat}, {Aulanier} \&
  {Schrijver}}]{masson09}
{Masson}, S., {Pariat}, E., {Aulanier}, G. \& {Schrijver}, C.~J. 2009 {The
  Nature of Flare Ribbons in Coronal Null-Point Topology}.
\newblock \emph{Astrophys. J.}, \textbf{700}, 559--578.
\newblock (\doi{10.1088/0004-637X/700/1/559})

\bibitem[{{McLaughlin} \& {Hood}(2004)}]{mclaughlin04}
{McLaughlin}, J.~A. \& {Hood}, A.~W. 2004 {MHD wave propagation in the
  neighbourhood of a two-dimensional null point}.
\newblock \emph{Astron. Astrophys.}, \textbf{420}, 1129--1140.
\newblock (\doi{10.1051/0004-6361:20035900})

\bibitem[{{McLaughlin} \& {Hood}(2006)}]{mclaughlin06}
{McLaughlin}, J.~A. \& {Hood}, A.~W. 2006 {MHD mode coupling in the
  neighbourhood of a 2D null point}.
\newblock \emph{Astron. Astrophys.}, \textbf{459}, 641--649.
\newblock (\doi{10.1051/0004-6361:20065558})

\bibitem[{{McLaughlin} \emph{et~al.}(2011){McLaughlin}, {Hood} \& {de
  Moortel}}]{mclaughlin11}
{McLaughlin}, J.~A., {Hood}, A.~W. \& {de Moortel}, I. 2011 {Review Article:
  MHD Wave Propagation Near Coronal Null Points of Magnetic Fields}.
\newblock \emph{Space Sci. Rev.}, \textbf{158}, 205--236.
\newblock (\doi{10.1007/s11214-010-9654-y})

\bibitem[{{Narukage} \emph{et~al.}(2014){Narukage}, {Shimojo} \&
  {Sakao}}]{narukage14}
{Narukage}, N., {Shimojo}, M. \& {Sakao}, T. 2014 {Evidence of Electron
  Acceleration around the Reconnection X-point in a Solar Flare}.
\newblock \emph{Astrophys. J. Lett.}, \textbf{787}, 125.
\newblock (\doi{10.1088/0004-637X/787/2/125})

\bibitem[{{Ofman}(2010)}]{ofman10}
{Ofman}, L. 2010 {Wave Modeling of the Solar Wind}.
\newblock \emph{Living Reviews in Solar Physics}, \textbf{7}, 4.
\newblock (\doi{10.12942/lrsp-2010-4})

\bibitem[{{Pariat} \emph{et~al.}(2009){Pariat}, {Antiochos} \&
  {DeVore}}]{pariat09}
{Pariat}, E., {Antiochos}, S.~K. \& {DeVore}, C.~R. 2009 {A Model for Solar
  Polar Jets}.
\newblock \emph{Astrophys. J.}, \textbf{691}, 61--74.
\newblock (\doi{10.1088/0004-637X/691/1/61})

\bibitem[{{Parnell} \& {De Moortel}(2012)}]{parnelldemoortel12}
{Parnell}, C.~E. \& {De Moortel}, I. 2012 {A contemporary view of coronal
  heating}.
\newblock \emph{Royal Society of London Philosophical Transactions Series A},
  \textbf{370}, 3217--3240.
\newblock (\doi{10.1098/rsta.2012.0113})

\bibitem[{{Parnell} \emph{et~al.}(2008){Parnell}, {Haynes} \&
  {Galsgaard}}]{parnell08}
{Parnell}, C.~E., {Haynes}, A.~L. \& {Galsgaard}, K. 2008 {Recursive
  Reconnection and Magnetic Skeletons}.
\newblock \emph{Astrophys. J.}, \textbf{675}, 1656--1665.
\newblock (\doi{10.1086/527532})

\bibitem[{{Parnell} \emph{et~al.}(2010{\natexlab{\emph{a}}}){Parnell}, {Haynes}
  \& {Galsgaard}}]{parnell10jgr}
{Parnell}, C.~E., {Haynes}, A.~L. \& {Galsgaard}, K. 2010{\natexlab{\emph{a}}}
  {Structure of magnetic separators and separator reconnection}.
\newblock \emph{J. Geophys. Res.}, \textbf{115}, A02102.
\newblock (\doi{10.1029/2009JA014557})

\bibitem[{{Parnell} \emph{et~al.}(2010{\natexlab{\emph{b}}}){Parnell},
  {Maclean} \& {Haynes}}]{parnell10apj}
{Parnell}, C.~E., {Maclean}, R.~C. \& {Haynes}, A.~L. 2010{\natexlab{\emph{b}}}
  {The Detection of Numerous Magnetic Separators in a Three-Dimensional
  Magnetohydrodynamic Model of Solar Emerging Flux}.
\newblock \emph{Astrophys. J. Lett.}, \textbf{725}, L214--L218.
\newblock (\doi{10.1088/2041-8205/725/2/L214})

\bibitem[{{Parnell} \emph{et~al.}(1994){Parnell}, {Priest} \&
  {Golub}}]{parnell94}
{Parnell}, C.~E., {Priest}, E.~R. \& {Golub}, L. 1994 {The three-dimensional
  structures of X-ray bright points}.
\newblock \emph{Solar Phys.}, \textbf{151}, 57--74.
\newblock (\doi{10.1007/BF00654082})

\bibitem[{{Parnell} \emph{et~al.}(1996){Parnell}, {Smith}, {Neukirch} \&
  {Priest}}]{Parnell96}
{Parnell}, C.~E., {Smith}, J.~M., {Neukirch}, T. \& {Priest}, E.~R. 1996 {The
  structure of three-dimensional magnetic neutral points}.
\newblock \emph{Phys. Plasmas}, \textbf{3}, 759--770.
\newblock (\doi{10.1063/1.871810})

\bibitem[{{Platten} \emph{et~al.}(2014){Platten}, {Parnell}, {Haynes}, {Priest}
  \& {Mackay}}]{platten14}
{Platten}, S.~J., {Parnell}, C.~E., {Haynes}, A.~L., {Priest}, E.~R. \&
  {Mackay}, D.~H. 2014 {The solar cycle variation of topological structures in
  the global solar corona}.
\newblock \emph{Astron. Astrophys.}, \textbf{565}, A44.
\newblock (\doi{10.1051/0004-6361/201323048})

\bibitem[{{Pontin} \emph{et~al.}(2013){Pontin}, {Priest} \&
  {Galsgaard}}]{pontin13}
{Pontin}, D.~I., {Priest}, E.~R. \& {Galsgaard}, K. 2013 {On the Nature of
  Reconnection at a Solar Coronal Null Point above a Separatrix Dome}.
\newblock \emph{Astrophys. J.}, \textbf{774}, 154.
\newblock (\doi{10.1088/0004-637X/774/2/154})

\bibitem[{{Pontin} \emph{et~al.}(2011){Pontin}, {Wilmot-Smith}, {Hornig} \&
  {Galsgaard}}]{pontin11}
{Pontin}, D.~I., {Wilmot-Smith}, A.~L., {Hornig}, G. \& {Galsgaard}, K. 2011
  {Dynamics of braided coronal loops. II. Cascade to multiple small-scale
  reconnection events}.
\newblock \emph{Astron. Astrophys.}, \textbf{525}, A57+.
\newblock (\doi{10.1051/0004-6361/201014544})

\bibitem[{{Priest} \& {Forbes}(2000)}]{priest2000}
{Priest}, E. \& {Forbes}, T. 2000 \emph{{Magnetic Reconnection}}.
\newblock Cambridge University Press, Cambridge, UK.

\bibitem[{{Priest} \& {D{\'e}moulin}(1995)}]{priest95}
{Priest}, E.~R. \& {D{\'e}moulin}, P. 1995 {Three-dimensional magnetic
  reconnection without null points. 1. Basic theory of magnetic flipping}.
\newblock \emph{J. Geophys. Res.}, \textbf{100}, 23\,443--23\,464.
\newblock (\doi{10.1029/95JA02740})

\bibitem[{{Priest} \& {Pontin}(2009)}]{priest09}
{Priest}, E.~R. \& {Pontin}, D.~I. 2009 {Three-dimensional null point
  reconnection regimes}.
\newblock \emph{Phys. Plasmas}, \textbf{16}(12), 122\,101.
\newblock (\doi{10.1063/1.3257901})

\bibitem[{{R{\'e}gnier} \emph{et~al.}(2008){R{\'e}gnier}, {Parnell} \&
  {Haynes}}]{regnier08}
{R{\'e}gnier}, S., {Parnell}, C.~E. \& {Haynes}, A.~L. 2008 {A new view of
  quiet-Sun topology from Hinode/SOT}.
\newblock \emph{Astron. Astrophys.}, \textbf{484}, L47--L50.
\newblock (\doi{10.1051/0004-6361:200809826})

\bibitem[{{Restante}(2011)}]{RestantePhD2011}
{Restante}, A.~L. 2011 {The Investigation of Quasi-Separatrix Layers in Solar
  Magnetic Fields}.
\newblock Ph.D. thesis, University of St Andrews.

\bibitem[{{Schindler} \emph{et~al.}(1988){Schindler}, {Hesse} \&
  {Birn}}]{schindler88}
{Schindler}, K., {Hesse}, M. \& {Birn}, J. 1988 {General magnetic reconnection,
  parallel electric fields, and helicity}.
\newblock \emph{J. Geophys. Res.}, \textbf{93}, 5547--5557.
\newblock (\doi{10.1029/JA093iA06p05547})

\bibitem[{{Stevenson} \& {Parnell}(2015)}]{Stevenson15b}
{Stevenson}, J.~E.~H. \& {Parnell}, C.~E. 2015 {Spontaneous Reconnection at a
  Separator Current Layer}.
\newblock \emph{Phys. Plasmas.}, \textbf{submitted}.

\bibitem[{{Stevenson} \emph{et~al.}(2015){Stevenson}, {Parnell}, {Priest} \&
  {Haynes}}]{Stevenson15}
{Stevenson}, J.~E.~H., {Parnell}, C.~E., {Priest}, E.~R. \& {Haynes}, A.~L.
  2015 {The nature of separator current layers in MHS equilibria. I. Current
  parallel to the separator}.
\newblock \emph{Astron. Astrophys.}, \textbf{573}, A44.
\newblock (\doi{10.1051/0004-6361/201424348})

\bibitem[{{Threlfall} \emph{et~al.}(2012){Threlfall}, {Parnell}, {De Moortel},
  {McClements} \& {Arber}}]{threlfall12}
{Threlfall}, J., {Parnell}, C.~E., {De Moortel}, I., {McClements}, K.~G. \&
  {Arber}, T.~D. 2012 {Nonlinear wave propagation and reconnection at magnetic
  X-points in the Hall MHD regime}.
\newblock \emph{Astron. Astrophys.}, \textbf{544}, A24.
\newblock (\doi{10.1051/0004-6361/201219098})

\bibitem[{{Threlfall}(2012)}]{threlfallPhD12}
{Threlfall}, J.~W. 2012 {Wave propagation, phase mixing and dissipation in Hall
  MHD}.
\newblock Ph.D. thesis, University of St.~Andrews (United Kingdom.

\bibitem[{{Thurgood} \& {McLaughlin}(2012)}]{thurgood12}
{Thurgood}, J.~O. \& {McLaughlin}, J.~A. 2012 {Linear and nonlinear MHD mode
  coupling of the fast magnetoacoustic wave about a 3D magnetic null point}.
\newblock \emph{Astron. Astrophys.f}, \textbf{545}, A9.
\newblock (\doi{10.1051/0004-6361/201219850})

\bibitem[{{Thurgood} \& {McLaughlin}(2013{\natexlab{\emph{a}}})}]{thurgood13b}
{Thurgood}, J.~O. \& {McLaughlin}, J.~A. 2013{\natexlab{\emph{a}}} {3D
  Alfv{\'e}n wave behaviour about proper and improper magnetic null points}.
\newblock \emph{Astron. Astrophys.}, \textbf{558}, A127.
\newblock (\doi{10.1051/0004-6361/201322021})

\bibitem[{{Thurgood} \& {McLaughlin}(2013{\natexlab{\emph{b}}})}]{thurgood13a}
{Thurgood}, J.~O. \& {McLaughlin}, J.~A. 2013{\natexlab{\emph{b}}} {Nonlinear
  Alfv{\'e}n wave dynamics at a 2D magnetic null point: ponderomotive force}.
\newblock \emph{Astron. Astrophys.}, \textbf{555}, A86.
\newblock (\doi{10.1051/0004-6361/201321338})

\bibitem[{{Titov}(2007)}]{titov07}
{Titov}, V.~S. 2007 {Generalized Squashing Factors for Covariant Description of
  Magnetic Connectivity in the Solar Corona}.
\newblock \emph{Astrophys. J.}, \textbf{660}, 863--873.
\newblock (\doi{10.1086/512671})

\bibitem[{{Titov} \emph{et~al.}(2003){Titov}, {Galsgaard} \&
  {Neukirch}}]{titov03}
{Titov}, V.~S., {Galsgaard}, K. \& {Neukirch}, T. 2003 {Magnetic Pinching of
  Hyperbolic Flux Tubes. I. Basic Estimations}.
\newblock \emph{Astrophys. J.}, \textbf{582}, 1172--1189.
\newblock (\doi{10.1086/344799})

\bibitem[{{Titov} \emph{et~al.}(2002){Titov}, {Hornig} \&
  {D{\'e}moulin}}]{titov02}
{Titov}, V.~S., {Hornig}, G. \& {D{\'e}moulin}, P. 2002 {Theory of magnetic
  connectivity in the solar corona}.
\newblock \emph{J. Geophys. Res.}, \textbf{107}, 1164.
\newblock (\doi{10.1029/2001JA000278})

\bibitem[{{Titov} \emph{et~al.}(2011){Titov}, {Miki{\'c}}, {Linker}, {Lionello}
  \& {Antiochos}}]{titov11}
{Titov}, V.~S., {Miki{\'c}}, Z., {Linker}, J.~A., {Lionello}, R. \&
  {Antiochos}, S.~K. 2011 {Magnetic Topology of Coronal Hole Linkages}.
\newblock \emph{Astrophys. J.}, \textbf{731}, 111.
\newblock (\doi{10.1088/0004-637X/731/2/111})

\bibitem[{{Titov} \emph{et~al.}(1993){Titov}, {Priest} \& {Demoulin}}]{titov93}
{Titov}, V.~S., {Priest}, E.~R. \& {Demoulin}, P. 1993 {Conditions for the
  appearance of ''bald patches'' at the solar surface}.
\newblock \emph{Astron. Astrophys.}, \textbf{276}, 564.

\bibitem[{{Ugarte-Urra} \emph{et~al.}(2007){Ugarte-Urra}, {Warren} \&
  {Winebarger}}]{ugarte-urra07}
{Ugarte-Urra}, I., {Warren}, H.~P. \& {Winebarger}, A.~R. 2007 {The Magnetic
  Topology of Coronal Mass Ejection Sources}.
\newblock \emph{Astrophys. J.}, \textbf{662}, 1293--1301.
\newblock (\doi{10.1086/514814})

\bibitem[{{Wang} \emph{et~al.}(2014){Wang}, {Young} \& {Muglach}}]{wang14}
{Wang}, Y.-M., {Young}, P.~R. \& {Muglach}, K. 2014 {Evidence for Two Separate
  Heliospheric Current Sheets of Cylindrical Shape During Mid-2012}.
\newblock \emph{Astrophys. J.}, \textbf{780}, 103.
\newblock (\doi{10.1088/0004-637X/780/1/103})

\bibitem[{{Wilmot-Smith} \emph{et~al.}(2009){Wilmot-Smith}, {Hornig} \&
  {Pontin}}]{wilmot-smith09}
{Wilmot-Smith}, A.~L., {Hornig}, G. \& {Pontin}, D.~I. 2009 {Magnetic Braiding
  and Quasi-Separatrix Layers}.
\newblock \emph{Astrophys. J.}, \textbf{704}, 1288--1295.
\newblock (\doi{10.1088/0004-637X/704/2/1288})

\bibitem[{{Wilmot-Smith} \emph{et~al.}(2010){Wilmot-Smith}, {Pontin} \&
  {Hornig}}]{wilmot-smith10}
{Wilmot-Smith}, A.~L., {Pontin}, D.~I. \& {Hornig}, G. 2010 {Dynamics of
  braided coronal loops. I. Onset of magnetic reconnection}.
\newblock \emph{Astron. Astrophys.}, \textbf{516}, A5+.
\newblock (\doi{10.1051/0004-6361/201014041})

\bibitem[{{Wyper} \& {Pontin}(2014)}]{wyper14}
{Wyper}, P.~F. \& {Pontin}, D.~I. 2014 {Non-linear tearing of 3D null point
  current sheets}.
\newblock \emph{Phys. Plasmas}, \textbf{21}(8), 082114.
\newblock (\doi{10.1063/1.4893149})

\end{thebibliography}

\end{document}